\documentclass[11pt,preprint]{aastex631}
\pdfoutput=1
\usepackage{bm}
\usepackage{graphicx}
\usepackage{footnote}
\usepackage{color}
\usepackage{amsmath}
\usepackage{amsfonts}
\usepackage{mathtools}
\usepackage{multirow}
\newcommand{\mubold}{\bm{\mu}}
\newcommand\ltsima{$\; \buildrel <\over\sim \;$}
\newcommand\simlt{\lower.5ex\hbox{\ltsima}}
\newcommand\gtsima{$\; \buildrel >\over\sim \;$}
\newcommand\simgt{\lower.5ex\hbox{\gtsima}}

\newcommand\msun {M_\odot}

\shorttitle{}
\shortauthors{Bhattacharya et al}

\begin{document}

\title{Confirmation of color dependent centroid shift measured after 1.8 years with HST}


\author{Aparna Bhattacharya}
\affiliation{Code 667, NASA Goddard Space Flight Center, Greenbelt, MD 20771, USA}
\affiliation{Department of Astronomy,
    University of Maryland, College Park, MD 20742, USA}
\email{abhatta5@umd.edu}

\author{David~P. Bennett}
\affiliation{Code 667, NASA Goddard Space Flight Center, Greenbelt, MD 20771, USA}
\affiliation{Department of Astronomy,
    University of Maryland, College Park, MD 20742, USA}

\author{Jean~Philippe~Beaulieu}
\affiliation{UPMC-CNRS, UMR 7095, Institut d’Astrophysique de Paris, 98Bis Boulevard Arago, F-75014 Paris}
\affiliation{School of Physical Sciences, University of Tasmania, Private Bag 37 Hobart, Tasmania 7001 Australia} 

\author{Ian~A.~Bond}
\affiliation{Institute of Natural and Mathematical Sciences, Massey University, Auckland 0745, New Zealand}

\author{Naoki~Koshimoto}
\affiliation{Department of Astronomy, The University of Tokyo, 7-3-1 Hongo, Bunkyo-ku, Tokyo 113-0033, Japan}
\affiliation{National Astronomical Observatory of Japan, 2-21-1 Osawa, Mitaka, Tokyo 181-8588, Japan}

\author{Jessica~R.~Lu}
\affiliation{University of California Berkeley, Berkeley, CA}

\author{Joshua~W.~Blackman}
\affiliation{School of Physical Sciences, University of Tasmania, Private Bag 37 Hobart, Tasmania 7001 Australia}

\author{Cl{\'e}ment~Ranc} 
\affiliation{UPMC-CNRS, UMR 7095, Institut d’Astrophysique de Paris, 98Bis Boulevard Arago, F-75014 Paris}

\author{Aikaterini Vandorou}
\affiliation{Code 667, NASA Goddard Space Flight Center, Greenbelt, MD 20771, USA}
\affiliation{Department of Astronomy,
    University of Maryland, College Park, MD 20742, USA} 
 
\author{Sean K. Terry} 
\affiliation{University of California Berkeley, Berkeley, CA}
 
\author{Jean~Baptiste~Marquette}
\affiliation{UPMC-CNRS, UMR 7095, Institut d’Astrophysique de Paris, 98Bis Boulevard Arago, F-75014 Paris}
 
\author{Andrew~A.~Cole}
\affiliation{School of Physical Sciences, University of Tasmania, Private Bag 37 Hobart, Tasmania 7001 Australia}
 
\author{Akihiko~Fukui}
\affiliation{Department of Earth and Planetary Science, Graduate School of Science, The University of Tokyo, 7-3-1 Hongo, Bunkyo-ku, Tokyo 113-0033, Japan}
\affiliation{Instituto de Astrof\'isica de Canarias, V\'ia L\'actea s/n, E-38205 La Laguna, Tenerife, Spain}

\begin{abstract}
%We present Keck/NIRC2 adaptive optics imaging of planetary microlensing event MOA-2007-BLG-400 that
%resolves the lens star system from the source. We find that the MOA-2007-BLG-400L planetary system consists
%of a $1.71\pm 0.27 M_{\rm Jup}$ planet orbiting a $0.69\pm 0.04\msun$ K-dwarf host star at a distance 
%of $6.89\pm 0.77\,$kpc from the Sun. So, this planetary system probably resides in the Galactic bulge. The 
%planet-host star projected separation is only weakly constrained due to the close-wide light curve degeneracy;
%the 2$\sigma$ projected separation ranges are 0.6--$1.0\,$AU and 4.7--$7.7\,$AU for close and wide solutions 
%respectively. This host mass is at the top end of the range of
%masses predicted by a standard Bayesian analysis. Our Keck follow-up program has now measured lens-source separations for six planetary 
%microlensing events, and five of these six events have host star masses above the median prediction
%under the assumption that assumes that all stars have an equal chance of hosting planets that can
%be detected by microlensing. This suggests that
%more massive stars may be more likely to host planets of a fixed mass ratio
%that orbit near or beyond the snow line. These results also indicate the importance of host star mass 
%measurements for exoplanets found by
%microlensing. The microlensing survey imaging data from NASA's {\it Nancy Grace Roman Space Telescope} 
%(formerly WFIRST) mission will be doing mass measurements like this for a huge number of planetary events. 
We measured precise masses of the host and planet in OGLE-2003-BLG-235 system, when the lens and source were resolving, with 2018 Keck high resolution images. This measurement is in agreement with the observation taken in 2005 with the Hubble Space Telescope (HST). In 2005 data, the lens and sources were not resolved and the measurement was made using color-dependent centroid shift only. Nancy Grace Roman Space Telescope will measure masses using data typically taken within 3-4 years of the peak of the event which is much shorter baseline compared to most of the mass measurements to date. Hence, color dependent centroid shift will be one of the primary method of mass measurements for Roman. Yet, mass measurements of only two events (OGLE-2003-BLG-235 and OGLE-2005-BLG-071) are done using the color dependent centroid shift method so far. The accuracy of the measurements using this method are neither completely known nor well studied. The agreement of Keck and HST results, shown in this paper, is very important since this agreement confirms the accuracy of the mass measurements determined at a small lens-source separation using the color dependent centroid shift method. This also shows that with $>$100 high resolution images, Roman telescope will be able to use color dependent centroid shift at 3-4 years time baseline and produce mass measurements. We find that OGLE-2003-BLG-235 is a planetary system consists
of a $2.34\pm 0.43 M_{\rm Jup}$ planet orbiting a $0.56\pm 0.06\msun$ K-dwarf host star at a distance 
of $5.26\pm 0.71\,$kpc from the Sun.    
   
\end{abstract}

\section{Introduction}
\label{sec-intro}
Gravitational microlensing is unique in its ability to detect low mass exoplanets \citep{bennett96} beyond the 
snow line \citep{gouldloeb1992}, where the formation of giant planets is thought to be most efficient
\citep{pollack96,lissauer_araa}. Roman Galactic Exoplanet Survey (RGES) was selected in 2010 decadal survey \citep{mpf,WFIRST_AFTA} to utilize this uniqueness and measure exoplanet demographics beyond the reach of Kepler mission. The aim of this survey is to detect low mass planets beyond the snowline of their hosts and measure the masses of the hosts and the planets. The high resolution capability of Roman telescope will measure masses of both the hosts and the planets. Currently, as part of a precursor study for RGES, we are developing the mass measurement methods (e.g. software for color dependent centroid shifts) for the exoplanets that are detected from the ground.   

OGLE-2003-BLG-235Lb is the first microlensing exoplanet discovered \citep{bond03} as well as the first event with mass measurements of the host star and the planet \citep{bennett06}. The lens host star is extremely faint when the source is magnified at its peak. Hence, the lens cannot be detected during this peak time of the microlensing event. However, years later, the source and lens have moved away from each other and the source is not magnified anymore. At this point, these two stars can be detected in high angular resolution images. If the source and lens are partially or fully resolved in the image, we observe the two stars appearing elongated and we call this image-elongation. We can detect the two stars in the image-elongation method easily and measure their masses. If the lens and source are not resolved but they have very different colors, they can still be detected in the high resolution images. For example, if the lens is redder than the source, the centroid of the flux of these two stars will shift from being closer to the lens in the red passband to being closer to the source in the blue passband. By measuring this centroid shift based on the color differences of the stars  (which is called the color-dependent centroid shift), we can detect the lens and measure the masses of the host and its planet. As \citet{bennett07} points out, the S/N of the image elongation method scales as $(\vartriangle x)^{2}$, whereas the S/N of the color dependent centroid shift scales as $(\vartriangle x)$. $(\vartriangle x)$ is the lens-source separation in this case. Hence, as long as the colors of the source and lens stars differ significantly, this method should measure small separation values more precisely than the image elongation method. That is the main advantage of the color dependent centroid shift method over image elongation method.  

 \citet{bennett06} observed the event with HST {\it Advanced Camera for Surveys} (ACS) instrument only 1.78 years after the peak of the event, which was about July 30, 2003. The event was observed in three passbands $B$, $V$ and $I$ to measure the color dependent centroid shifts between three different passbands. The foreground lens (host star) is brighter in red passbands and fainter in blue passbands, this causes the centroid of the lens-source flux to shift between red to blue passbands. The measurement of this centroid shift enabled the measurement of the microlens host and planet masses for the first time. 

 The mass measurement we present for this study has the second shortest time gap between the peak of the event and the follow up observation. OGLE-2005-BLG-071, with 0.84 years gap, has the shortest time gap between the peak of the event and the follow-up observations \citep{bennett20}. In the current paper, we confirm the 2006 color dependent centroid shift results and the mass measurements with our 2018 Keck observations. We observe the event with Keck NIRC2 Adaptive Optics (AO) in 2018 and detect the lens host star separated from the source. We measure the lens-source relative proper motion and  predict the color dependent centroid shift in 2005 data and compare with the \citet{bennett06} work. We also measure the lens properties independently from 2018 Keck data and use them to confirm the masses obtained from the 2006 paper. This work is important since this will be one of the primary methods of measuring the mass of the host stars and the planets with the      
 {\it Nancy Grace Roman Space Telescope} ( hereafter, {\it Roman} telescope)  (formerly WFIRST) mission,
which is NASA's next astrophysics flagship mission, to follow the {\it James Webb Space Telescope} (JWST).

The {\it Nancy Grace Roman Space Telescope} \citep{WFIRST_AFTA} includes the {\it Roman Galactic Exoplanet Survey} (RGES)\citep{penny19} , 
that will detect and measure the mass of hundreds of exoplanets in orbits 
extending from the habitable zone to infinity (i.e.\ unbound planets). The microlens exoplanets discovered by
{\it Roman} will not require follow-up observations since the {\it Roman} observations will also have 
high enough angular resolution to detect the lens (and planetary host) stars \citep{bennett07}. The RGES will observe the galactic bulge in six to eight epochs across the 5 year time-span. Hence the maximum time separation between two epochs will be about $\sim$ 4.5 years. This is because, in spite of the prime mission being 5 years long, the time difference between the first and the last galactic bulge season will be about 4.5 years. For example, a planetary event detected in the first epoch, will be going through lens-source separation in the final epoch. So the masses of this system can be measured using the centroid shift method. However, this color dependent centroid shift method for small separations of the lens and the source is not confirmed except for one event \citep{bennett20}. Very few color dependent centroid shift measurements are done, some of them with more than 4.5 years after the peak of the event. Hence, confirming such a color dependent centroid shift measurement, especially from such a small time separation, is important for demonstrating that RGES can properly conduct mass measurements.  

 This event is observed as part of our  
NASA Keck Key Strategic Mission
Support (KSMS) program whose main purpose is to refine the mass measurement method, and 
optimize the {\it Roman} exoplanet microlensing survey observing program. We are conducting a systematic exoplanet microlensing event high angular resolution follow-up program
to detect and determine the masses of the exoplanet host stars with our NASA  Keck KSMS program \citep{bennett_KSMS}. This is supplemented by HST observations \citep{aparna19hst}
for host stars that are most likely to be detected with the color dependent centroid shift method
\citep{bennett06}. This program has already revealed a number of microlens exoplanet host stars that are 
resolved from the source stars \citep{van20,bennett20,blackman21,terry21}, and others that are still blended
with their source stars, but show a significant elongation due to a lens-source separation
somewhat smaller than the size of the point spread function (PSF) \citep{bennett07,ogle169,aparna18}. This program has also recently provided mass upper limits of a Jovian planet orbiting a White Dwarf host. Another purpose of our NASA KSMS program is to confirm measurements of the small lens-source separation with undersampled pixels ( eg. \citet{bennett20}. In our current paper we will confirm the small lens-source separation measured by under-sampled HST pixels with Keck high angular resolution images. 

The paper is organized as follows: Section \ref{sec-model} revisits the ground based seeing limited photometry data from 2003 and re-analyzes the light curve modeling. The main purpose of this re-analysis is to obtain the MCMC chains that are necessary for calculating the uncertainties in the mass measurements. The light curve modeling from \citep{bond03} did not provide these MCMC chains. Section \ref{sec-Followup} describes the details of our high resolution follow up observations and the reduction of the AO images. In the next two sections, we show the process of identifying 
the host star (which is also the lens) and the source star. In section \ref{sec-centroidshift}, 
we predict the color dependent centroid shifts between $B$, $V$ and $I$ passbands in 2005 data based on the geocentric relative lens-source proper motion from the lens and source identification in 2018 Keck data. Note that, from here on, we address the results of \citet{bennett06} as analysis of 2006 paper though it refers to the analysis of 2005 Hubble data. We 
show that this prediction matches with the centroid shift measurements in \citet{bennett06} paper. Finally in sections \ref{sec-lens} and \ref{sec-discussion}, 
we discuss the exoplanet system properties and the implications of its mass and distance measurement.    

\begin{deluxetable}{cc}
\tablecaption{Best Fit Model Parameters
                         \label{tab-mparams} }
\tablewidth{0pt}
\tablehead{
%% Use a footnote to explain numbering.
%& & & & \multicolumn{2}{c} {MCMC averages} \\
\colhead{parameter}  & \colhead{Values} }  
  % end header.
\startdata
%                      pleb_2    plmb_2 
$t_E$ (days) & $61.2 \pm 2.2$     \\   
$t_0$ (${\rm HJD}^\prime$) & $2848.04 \pm 0.11$  \\
$u_0$ & $-0.13 \pm 0.61$   \\
$s$ & $1.12 \pm 0.43$  \\
$\alpha$ (rad) & $0.76 \pm 0.15$   \\
$q \times 10^{-3}$ & $3.99 \pm 0.48$   \\
$t_\ast$ (days) & $0.061 \pm 0.004$  \\
$\theta_{E}$ (mas) & $0.59 \pm 0.07$\\
$V_S$ & $21.23 \pm 0.05$   \\
$I_S$ & $19.72 \pm 0.05$  \\
$\mu_{\rm rel,G}\,$(mas/yr) & $3.48 \pm 0.14$ \\
%$H_S$ & 16.663 & 16.645 &$16.657\pm 0.030$ \\
%fit $\chi^2$ & 3136.95 & 3136.86 \\
\enddata
\end{deluxetable}

\section{Revisiting Photometry and Light Curve Modeling}
\label{sec-model}

The light curve modeling of this event was first published in \citet{bond03}. However, the OGLE III data used for this modeling was not calibrated until 2010. Hence, in this paper, we are performing the remodeling of the light curve photometry with calibrated OGLE III $V$ data.   
This was done using the \citet{bennett-himag} modeling code,
and the resulting model parameters are given in Table~\ref{tab-mparams}.
The model parameters
that are in common with single lens events are the Einstein radius crossing time, $t_E$, the time, $t_0$, 
and distance, $u_0$, of closest alignment between the source and the lens center-of-mass, where $u_0$ is 
given in units of the Einstein radius. There are four additional parameters for binary lens systems: the
star-planet separation in Einstein radius units, $s$, the angle between the lens axis and the source 
trajectory, $\alpha$, the planet-star mass ratio, $q$, and the source radius crossing time, $t_*$, 
which is needed for events, like most planetary events, that have very sharp intrinsic light curve 
features that resolve the angular size of the source star. The brightness of the source star, $f_{Si}$ 
and blended stars, $f_{Bi}$ are also fit to the observed brightness for each passband, $i$, using the
formula $F_i(t) = f_{Si}A(t) + f_{Bi}$, where $A(t)$ is the magnification from the model, and $F_i(t)$
is the observed flux in the $i^{\textrm{th}}$ passband. Because this is a linear equation, $f_{Si}$ and $f_{Bi}$ can
be solved for each model in the Markov Chain \citep{rhie99}. 
%For each data set
%that has been calibrated, the $ f_{Si}$ values are used to determine the calibrated source brightness.

%Another improvement in our analysis is the measurement of the lens-source relative proper motion,
%$\mu_{\rm rel,G} = 8.79\pm 0.18\,$mas/yr. The G suffix refers to the use of the inertial
%Geocentric reference frame that moves with the
%velocity of the Earth at the time of the event. This new measurement compares to $\mu_{\rm rel,G} = 8.0\pm 0.46\,$mas/yr as reported in \citet{dong-moa400}. There are several ingredients to this improvement. The improved 
%CTIO photometry provides a more accurate color measurement, and the analysis of \citet{boyajian14} as
%optimized for microlensing targets \citep{aparna16} provides a more accurate source radius.
 Another improvement is that  
\citet{nataf13} has provided a more accurate determination of the properties of the red clump giants
that are used to determine the dust extinction in the foreground of the source. This has resulted in a more accurate source radius and a relative lens-source proper motion, $\mu_{\rm rel,G}$. The $\mu_{\rm rel,G}$ 
prediction is important because it can be used to confirm our planetary interpretation of the
light curve \citep{ogle169,batistaogle169}. 
%{\bf are these above sentences are still relevant for this event since the masses and planetary interpretation were already published. }

\section{Follow up observations}
\label{sec-Followup}

\subsection{Wide Camera}
The event was observed  with the Keck Adaptive Optics (AO) NIRC2 \citep{keckAO}
instrument during the  early 
morning of Aug 05, 2018. Five dithered exposures, 
each of 30 seconds, were taken in the $K_S$ short passband ( hereafter, $K$ band) with the wide camera. Each wide camera image covers a 1024 $\times$ 1024 square 
pixel area, and each pixel size is about $39.686 \times 39.686\,$mas. These images were flat field and dark current 
corrected using standard methods, and then stacked using the SWarp Astrometics package \citep{SWarp}. 
The details of our methods are described in \citet{batista2014}. We used aperture photometry method on these 
wide images with SExtractor software \citep{sextractor}. These wide images were used to detect 
and match 22 bright isolated stars to the VVV catalog  \citep{vvv} for calibration purposes. The 
average FWHM of this wide camera stack image is 90 mas.  Note that, in the 
2018 wide camera images, the lens and source were not resolved. 
As a result, we need NIRC2 narrow camera images to resolve and to identify the lens system.   
\subsection{Narrow Camera}

This event was also observed with the Keck NIRC2 narrow camera on Aug 05, 2018 in the $K$-band using laser guide star adaptive optics (LGSAO).  The 
main purpose of these images is to resolve the lens host star from the source star. Sixteen dithered observations were taken with 
60 seconds exposures. The images were taken with a small dither of 0.7'' at a position angle (P.A.) of 0$^{\circ}$ with each frame consisting of three co-added 20 seconds integrations. 
%The seeing of these narrow cameraimages was $\sim$0.4-0.6''. 
The average FWHM of these images is $\approx$50 mas.  For the reduction of these images, we used nine dark images of 60 seconds exposure time and 10 $K$-band dome flats 
taken with narrow camera on the same day as the science images. There were five dome flat  images with 
the lamp on and five more images with the lamp off, each with 60 seconds exposure time. Also at the end of 
the night, we took 31 sky images using a clear patch of sky at a (RA, Dec) of (20:29:57.71, -28:59:30.01) 
with an exposure time  of 30 seconds each. All these images were taken with the $K$ band filter. These images were used to flat field, bias subtract and 
remove bad pixels and cosmic rays from the eleven raw science images. The strehl ratio of these clean images varied
over the range 0.21-0.41. Finally 
these clean raw images were distortion corrected, differential atmospheric refraction corrected and stacked into one image \citep{jlu_thesis, KAI}. We used the stacked image for the final photometry and
astrometry analysis.

% This is probably due to the fact that AO correction for faint stars are better in $K$ band than in $H$ band.  
% Your statement above is not correct. The AO corrections are the same for faint and bright stars.

There are 1024$\times$1024 pixels in each narrow camera image with each pixel subtending
9.942 mas \citep{refraction, distortion} on each side. Since the small field of these narrow images covers only a few bright stars, 
it is difficult to directly photometrically calibrate them to VVV. On the contrary, wide images cover 1024$\times$1024 pixels with each pixel subtending 39.686 mas on each side. So we use the wide camera $K$  stack image that 
was already calibrated to VVV (see above section 3.1) to calibrate the narrow camera images. The wide and narrow images were taken in same filter, hence we did not need to do any filter conversion.
This gives us the brightness calibration between the stacked narrow camera image to VVV image. The 
photometry used for the narrow camera image calibration is from DAOPHOT analysis (section \ref{sec-Keck}). 

\section{Keck Narrow Camera Image Analysis} 
\label{sec-Keck}
%\subsection{2018 Narrow Data}
In this section, we use DAOPHOT \citep{Daophot} to construct a proper empirical PSF model to identify the two stars (the lens and the source) in the narrow stack images. We started our analysis with 
the 2018 Keck narrow camera images. We used the same method as \citet{aparna18} to build PSF models for the narrow camera stack images. We built these 
PSF models in two stages. In the first 
stage, we ran the FIND and PHOT commands of DAOPHOT to find all the possible stars in the 
image. In second stage, we used the PICK command to build a list of bright ($K < 18.5$) isolated stars that can be 
used to construct our empirical PSF model. Our target object was excluded from this list of PSF stars because it
is expected to consist of two stars that are not in the same position. From this list, we selected the 4 nearest stars 
to the target that had sufficient brightness, and we built our PSF model from these stars. We chose only the 
nearest stars in order to avoid any effect of PSF shape variations across the image. 
%We used the same PSF stars for both $K$ and $H$ band data sets. 

\begin{figure}
\epsscale{1.0}
\plotone{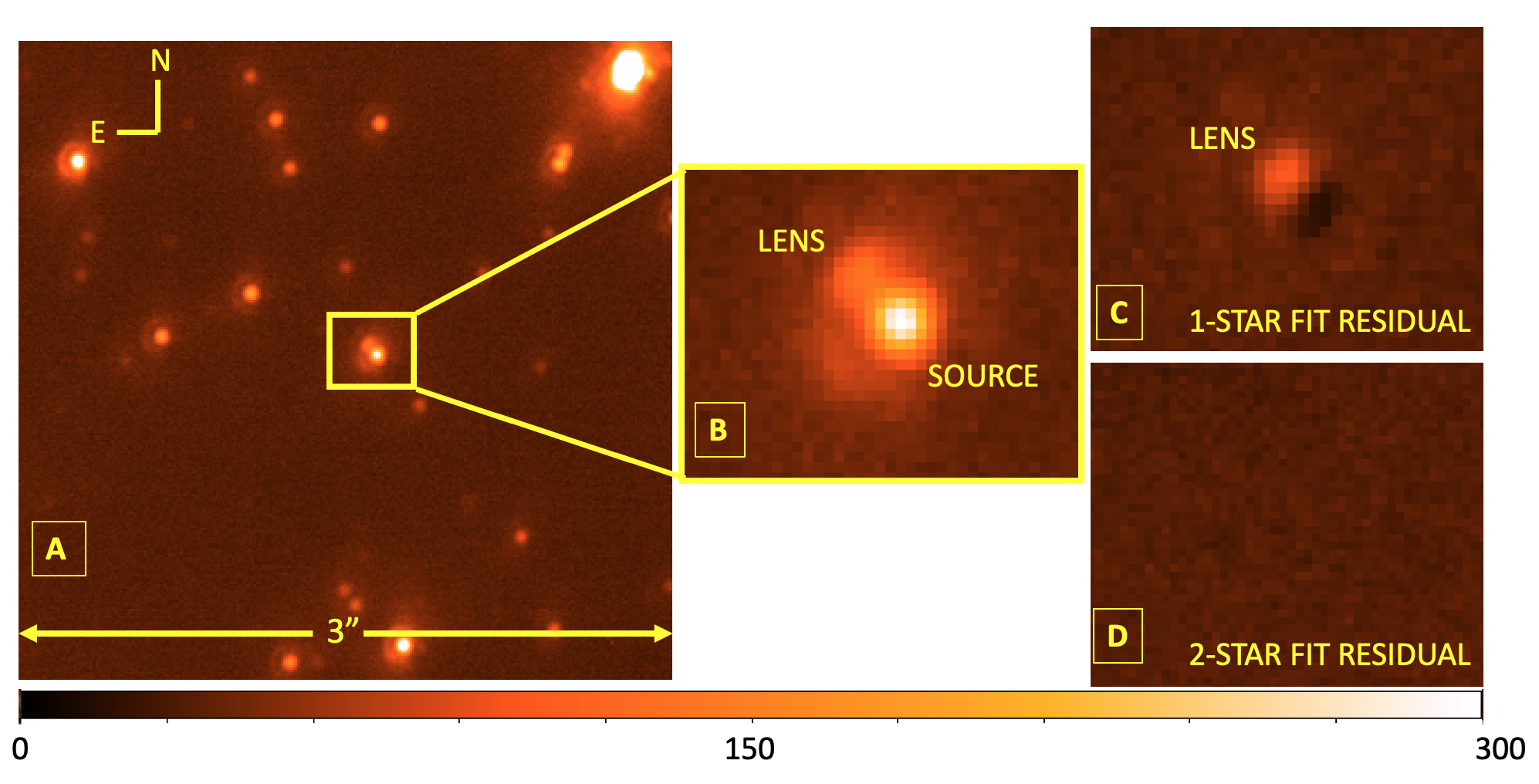}
\caption{{\it Panel A}: A 3" section of the stack image of 16 Keck $K$-band images, taken with the narrow camera,
with the yellow square indicating the target.
{\it Panel B}: A close-up of the target object. The source and lens 
positions are obtained from the best fit dual star PSF model. {\it Panel C}: The residual image after subtracting 
the best fit single star PSF model. The residual shows a clear indication of the presence of an additional star East of the target. {\it Panel D}: The residual image after 
subtracting the best fit dual star PSF model. This shows a smooth residual, confirming the second star at 
 separation of $\sim$54 mas. (C) and (D) use the same photometry scale.}
\label{fig-keck}
\end{figure}

 Once we have built the PSF model, we fit all the stars with this model.  
%we carefully checked the residual image that has
%all the identified stars subtracted. We noticed that the residuals of all the single stars looked different in different parts of the image. This indicates the PSF varies over the field, an effect that is particularly pronounced for the $H$ band stack image. So, we reconstructed our empirical PSF model using only 4 bright, isolated stars near the target object. This was our final PSF model. We ran the PSF fitting again on the field 
%with this new PSF model. 
This step produced the single star residual fit for the target that is shown in Figure \ref{fig-keck}C. Inspection of this residual image from the single star fit 
indicates that there is an additional star near the target object. So, we tried fitting the region of the target object with a dual star model. The dual star fits produced a smoother residual than the single star fit, as shown in Figure \ref{fig-keck}D. The results of these dual star fits are given in Tables \ref{tab-Keck_phot} and \ref{tab-murel}. Both the single star and dual star fits were done using the Newton-Raphson method of standard DAOPHOT. There is a minor wing-like feature to the south and east of the source star in Figure 1B. However, this feature is seen in all the nearby stars in the image and have been easily taken care of by the empirical PSF that is built from the nearby star brightness profiles. Hence in Figures 1C and 1D, this feature is subtracted clearly in the residual images. Keck PSFs are often not perfectly round due to the instability of atmosphere. However, the method of empirical PSF building from the nearby stars seem to compensate for that and do a proper PSF fitting even when the PSFs are not perfectly round. 
   
As we discuss in Section~\ref{sec-keck-inter}, we identify the brighter of the two stars as the source for the
OGLE-2003-BLG-235 microlensing event, and the fainter star to the North-East as the lens and planetary host star.

%We noticed that the residuals of the dual star fit in both passbands are not as clean as some of our previous analyses \citep{batistaogle169, aparna18,van20}. However, most of the single bright stars in the image have similar residual as the final two star fits. Hence, a possible reason for this could be the PSF model built from fainter stars is not sufficiently good for the bright star. A more robust way to deal with this problem is to build separate PSF models for the bright star and the faint star. {\bf [This discussion is strange. Generally, the PSF shape should not depend on brightness, unless you are close to saturation where the detector response can become non-linear. A more plausible reason for the larger residuals for brighter stars is simply that they are brighter so that their residuals are above the noise. A more likely scenario is simply that DAOPHOT has failed to construct an imperfect PSF model, which leaves significant residuals for the brighter stars.]} However, in the current version of DAOPHOT, there is no way to do dual star fits with two different PSF models simultaneously.   

%\subsection{Error Bars from the Jackknife Method \label{sec-jackknife}}
We have calculated the error bars using 
the Jackknife method \citep{quenouille1949,quenouille1956, tukey1958, tierney} following \citet{aparna21}. This method is 
able to measure the variance due to the 
PSF variations in the individual images. There were a total of 16 clean Keck images. In this method, we built 16 different stacks of 15 images, by removing one of the 16 images each time. 
Then, these 16 images are analyzed with DAOPHOT to build empirical PSFs for each stack image. Next, the target in each image is fit with the dual star PSF models. Once we have 16 sets of positions and flux of lens and source, the sample mean and the standard deviation of the parameters are used for the statistics. The standard error of a parameter $x$ in Jackknife is given by: 
\begin{equation}
\label{eq-jackknife}
SE(x) = \sqrt{\frac{N-1}{N} \sum (x_{i} - \bar{x})^{2}} \ .
\end{equation}  
The $x_{i}$ represents the value of the parameter measured in each of the combined image and $\bar{x}$ represents the mean of the parameter $x$ from all the N stacked images. This Equation \ref{eq-jackknife} is the same 
formula as the sample mean error, except that it is multiplied by $\sqrt{N-1}$. These error bars are reported in Table \ref{tab-Keck_phot}. The error bars in Table \ref{tab-murel} 
are based on the error bars in Table \ref{tab-Keck_phot} because the Heliocentric lens-source relative proper motion, 
$\mu_{\rm rel,H}$, is proportional to the lens-source separation.

\begin{deluxetable}{ccccc}
\tablecaption{Source and Lens Photometry and Astrometry from 2018 Keck data\label{tab-Keck_phot}}
\tablewidth{0pt}
\tablehead{ \multicolumn{2}{c}{Calibrated Magnitudes}&\multicolumn{3}{c}{Lens- Source Separation(mas)} \\
Source &Lens& East& North& total}
\startdata
$17.76 \pm 0.06$ & $19.20 \pm 0.11$ & $37.07 \pm 0.18$ & $38.57 \pm 0.45$ & $53.5 \pm 0.48$ \\
\enddata
%\vspace{0.15cm}
\tablecomments{The separation was measured 15.01 years after the peak of the event in the 
$K$ band. The follow-up observation was taken on Aug 05, 2018 and the peak of the microlensing light curve event was on July 30, 2003.}   
% We should report the time interval to better than 1% precision because the measurement is better than 2%.
\end{deluxetable}

\section{Interpreting Keck Results \label{sec-keck-inter}}     

In previous sections, we have referred to the Keck images of the source and lens (and planetary host) stars.
Now, we discuss the analysis that shows that our identifications of the source and lens stars are correct. Based on these identifications, we compare our 2018 Keck results to the 2006 \citep{bennett06} result and show that they are consistent in section \ref{sec-lens}.

\subsection{Confirming the Source Star Identification}
 Our reanalysis of the light curve model of this event 
(with improved OGLE III calibration) showed the best fit model gives the source brightness $I_S = 19.72 \pm {0.05}$ and $V_S = 21.23 \pm {0.05}$. The average extinction and reddening of the red clump stars within $2.5^\prime$ of the target are $A_I = 1.22$ and $E(V-I)= 0.92$, as calculated from the OGLE-III photometry catalog \citep{ogle3-phot}. 
This means the extinction corrected source magnitudes are $I_{S,0} = 18.49 \pm {0.06}$  and  $V_{S,0} = 19.09 \pm {0.06}$.
The extinction corrected color is $(V-I)_{S,0} = 0.60 \pm {0.04} $. From the color-color relation 
of \citet{kenyon_hartmann}, 
$(V-I)_{S,0}=0.60$ corresponds to the dereddened colors of $(V-K)_{S,0}=1.28$. 
So the extinction corrected source brightnesses in $K$ is $K_{S,0} = 17.81 \pm {0.06}$ including 
a 5$\%$ uncertainty in the color-color relation. The extinction 
in the $K$ band is $A_K = 0.13$ \citep{dutra_2003}, hence  
 the expected $K$ band 
calibrated magnitude of the source is $K_S = 17.94 \pm 0.08$.      

\begin{deluxetable}{ccccc}
\tablecaption{Measured Lens-Source Relative Proper Motion\label{tab-murel}}
\tablewidth{0pt}
\tablehead{
\multicolumn{3}{c}{$\mathbf{\mu}_{\rm rel,H}$(mas/yr)}&\multicolumn{2}{c}{$\mathbf{\mu}_{\rm rel,G}$(mas/yr)}\\
 $\bm{\mu}_{\rm rel,H,E}$&$\bm{\mu}_{\rm rel,H,N}$ & $|\mathbf{\mu}_{\rm rel,H}|$ &
 Measured from Keck & From light curve }
\startdata
   $2.47 \pm 0.01$ & $2.57 \pm 0.03$ & $3.56 \pm 0.08$ & $3.31 \pm 0.06$ & $3.48 \pm 0.15$  \\
% Keck $H$  & $8.43 \pm 0.23$ & $-2.56\pm 0.19$ & $1.79\pm 0.21$ & $-8.63\pm 0.22$ \\
%Mean &$8.52\pm0.13$ &$-2.29\pm 0.13$ &  $2.07\pm 0.13$ & $-8.55\pm 0.13$ \\
\enddata
\end{deluxetable}

 The dual star fits to 2018 narrow camera Keck images imply that the two stars located at the position of the target are resolved.
The best dual star fit yielded two stars with calibrated $K$ magnitudes of $17.76 \pm 0.06$ and $19.20 \pm 0.11$. 
 The error bars are 
calculated using the jackknife method, as discussed in section \ref{sec-Keck}.  The brighter star from 
these dual star fits matches the expected source magnitudes derived above. 
Hence, we identify the brighter star as the source star and the fainter star as the potential lens star.

\subsection{Determination of Relative Lens-Source Proper Motion}
\label{sec-murel}

 We will determine the relative proper motion between the lens candidate and the source star. In this subsection, we will measure the heliocentric relative proper motion from narrow Keck images and then transform that to geocentric frame. Then we will compare the measured geocentric relative proper motion to the predicted relative lens-source proper motion from the light curve. This will help us to confirm that the second star is indeed the lens star later. Since in the next section we show that this possible lens star is the lens, we will call the relative proper motion as relative lens-source proper motion from now for simplicity. 

The two stars are separated by slightly more than the FWHM of the $K$ band image,with the separations given in Table \ref{tab-Keck_phot}.
 The measured separations can be used to compute the lens-source relative proper motion, $\mu_{\rm rel}$, 
which can be compared with the $\mu_{\rm rel}$ prediction from the light curve. However, this issue is complicated
by the fact that the lens-source separation measurements determine the relative proper motion in a Heliocentric
frame, $\mu_{\rm rel,H}$, while the light curve measures the relative proper motion in an inertial Geocentric reference
frame, $\mu_{\rm rel,G}$, that moves with the Earth's velocity at the time of the event. The relationship between
these reference frames is given by equation \ref{eq-mu_helio}.

 Our high resolution observations of 2018 were taken $\sim$15.01 years after the microlensing event magnification peak. Since these images were taken almost exactly 15 years after the microlensing magnification, we do not need to consider the relative position of the Earth with respect to the Sun when determining the heliocentric lens-source relative proper motion
from the measured separations in the Keck images.

At the time of peak magnification, the separation between lens and source was 
$\sim |u_0\theta_E| \sim 0.017\,$mas. 
Hence, by dividing the measured separation by the time interval 
of 15.01 years, we obtain the heliocentric lens-source 
relative proper motion, $\mubold_{\rm rel,H}$. A comparison of these values from our independent dual star
fits for the $K$ band is shown in Table \ref{tab-murel} with 
error bars are estimated from jackknife method. 
%In Galactic coordinates, the mean $\mubold_{\rm rel,H}$ components are 
%$\mu_{{\rm rel,H},l} = 2.07 \pm 0.13$ mas/yr and  
%$\mu_{{\rm rel,H},b} = -8.55\pm 0.13$ mas/yr, with an amplitude
%of $\mu_{\rm rel,H} = 8.80\pm 0.18$ mas/yr at an angle of $\sim -74^\circ$ from the direction of Galactic disc rotation. 
%The dispersion in the motion of stars in the bar shaped bulge at the lens distance of 
%$D_L \approx 6.97\,$kpc (as presented in section \ref{sec-lens}) is about $\sim 2.5\,$mas/yr in each direction. 
%The source is also in the bulge at about $\sim 8\,$kpc, where a similar dispersion in the motion of stars is expected. 
%The relative proper motion is the difference of two proper motions, so the average difference in proper motion is
%the quadrature sum of four  $\sim 2.5\,$mas/yr values or $\sim 5\,$mas/yr. However the microlensing rate 
%is proportional to $\mu_{\rm rel}$, so the average $\mu_{\rm rel}$ is greater than $> 5\,$mas/yr. So, our
%measured $\mubold_{\rm rel,H}$ value is only slightly higher than the typical value for bulge-bulge lensing events. 

Our light curve models were done in a geocentric reference frame that differs from the heliocentric
frame by the instantaneous velocity of the Earth at the time of peak magnification, because the 
light curve parameters can be determined most precisely in this frame. However, this also means that
the lens-source relative proper motion that we measure with follow-up observations is not in the
same reference frame as the light curve parameters. This is an important issue because, as we
show below (see section \ref{sec-lens}), the measured relative proper motion can be combined with brightness of
the source star determine the mass of the lens system. The relation between the relative
proper motions in the heliocentric and geocentric coordinate systems are given by \citet{dong-moa400}:
%{\bf [\citep{dong-moa400} is the wrong reference. Find the right one.]}
\begin{equation}
\bm{\mu}_{\rm rel,H} = \bm{\mu}_{\rm rel,G} + \frac{{\bm v}_{\oplus} \pi_{\rm rel}}{\rm AU}  \ ,
\label{eq-mu_helio}
\end{equation}
where ${\bm v}_{\oplus}$ is the projected velocity of the earth relative to the sun (perpendicular to the 
line-of-sight) at the time of peak magnification. 

The projected velocity for OGLE-2003-BLG-235 is
${{\bm v}_{\oplus}}_{\rm E, N}$ = (24.53, -1.304) km/sec = (5.17, -0.28) AU/yr at the peak of the microlensing. The relative parallax is 
defined as $\pi_{\rm rel} \equiv  (1/D_L - 1/D_S)$, where $D_L$ and $D_S$ are lens and source distances. Hence the Equation \ref{eq-mu_helio} can written as:
\begin{equation}
\bm{\mu}_{\rm rel,G} = \bm{\mu}_{\rm rel,H} - (5.17, -0.28 )\times (1/D_L - 1/D_S) 
\label{eq-mu}
\end{equation}
Since $\bm{\mu}_{\rm rel,H}$ is already measured in Table \ref{tab-murel}, Equation 3 yields the geocentric relative proper motion, $\bm{\mu}_{\rm rel,G}$ as a function of the lens distance. Now at each possible lens distance, 
we can use the $\mu_{\rm rel,G}$ value from equation~\ref{eq-mu} to determine the angular 
Einstein radius, $\theta_E = \mu_{\rm rel,G} t_E$. Since we already know the $\theta_E$ value from the light curve models, we can use that here to constrain the lens distance and relative proper motion. Using this method, we determined the
relative proper motion to be $\mu_{\rm rel,G} = 3.31 \pm 0.06$. 
This value is consistent with the predicted $\mu_{\rm rel,G} = 3.48 \pm 0.15$ mas/yr from the light curve models.

\subsection{Confirmation of the Host Star Identification}
\label{sec-lens-id}

We can confirm our identification of the lens star only if the measured lens source relative proper motion is consistent with the predicted one from the light curve. Note that, the measured relative proper motion in Keck images is in Heliocentric frame. In order to compare the measured relative proper motion to the light curve prediction of $\mu_{\rm rel,G} = 3.48 \pm 0.15\,$mas/yr, we used 
equation~\ref{eq-mu_helio} to convert between the Geocentric and Heliocentric coordinate systems in section \ref{sec-murel}. This
requires knowledge of the source and lens distances, and we do this comparison inside our Bayesian 
analysis, presented in Section~\ref{sec-lens}, which combines the Markov chain of light curve models
with the constraints from the Keck observations and the Galactic models. As shown in Table \ref{tab-murel}, the measured and predicted $\mu_{\rm rel,G}$ are consistent within $1\sigma$. This shows that the faint star in Table \ref{tab-Keck_phot} is the lens. 

In addition, Table \ref{tab-params-histogram} shows the measured relative proper motion in 2006 HST analysis paper of this event. The fainter star was much closer to the source in 2005 data and at a much farther distance in 2018 data. The consistency between 2005 and 2018 measurements with the same lens-source relative motion shows that it is indeed this star that is moving away from the source. This proves that this faint star is the lens and not a nearby unrelated star.    

\section{Lens Properties}
\label{sec-lens}
Mass measurements of microlensing exoplanets and their hosts can be done solving a minimum of two of the three mass distance relations. 
Measurements of the angular Einstein radius, $\theta_E$, and the microlensing parallax 
amplitude, $\pi_E$, can each provide mass-distance relations \citep{bennett_rev,gaudi_araa},
\begin{equation}
M_L = {c^2\over 4G} \theta_E^2 {D_S D_L\over D_S - D_L} = {c^2\over 4G}{ {{\rm AU}^2}\over{\pi_E}^2}{D_S - D_L\over D_S  D_L}  \ .
\label{eq-m_thetaE}
\end{equation}
 $D_L$ and $D_S$ are distances to the lens and the source respectively. These can be combined to yield the lens mass in an expression with no dependence on the lens or source distance,
\begin{equation}
M_L = {c^2 \theta_E {\rm AU}\over 4G \pi_E} = {\theta_E \over (8.1439\,{\rm mas})\pi_E} \msun \ .
\label{eq-m}
\end{equation}
The angular Einstein radius can be measured for most planetary microlensing events because 
most planetary events have finite source effects that allow the measurement of the source radius crossing time,
$t_*$. The angular Einstein radius is then given by $\theta_E = \theta_* t_E/t_*$, where $t_E$ is the Einstein radius crossing time and
$\theta_*$ is the angular source radius, which can be determined from the
source brightness and color \citep{kervella_dwarf,boyajian14}. As a result, the measurement of $\pi_E$ for planetary events
usually results in mass measurements. Unfortunately, the orbital motion of the Earth allows $\pi_E$ 
to be determined for only a relatively small subset of events such as the ones that have very long durations
\citep[e.g.][]{gaudi-ogle109,bennett2010}, long duration events with bright source stars \citep[e.g.][]{muraki11},
and events with special lens geometries \citep[e.g.][]{sumi16}. The microlensing parallax program using the
{\it Spitzer} space telescope at $\sim 1\,$AU from Earth has recently expanded the number of 
events with microlensing parallax measurements \citep{udalski_ogle124,street16}, but recent studies
indicates that systematic errors in the {\it Spitzer} photometry can affect
some of the {\it Spitzer} $\pi_E$ measurements \citep{spitz_vs_gal,gould20,dang20}.

The method that can determine the masses of the largest number of planetary microlensing events is
the detection of the exoplanet host star as it separates from the background source star. However, due to the high stellar density in the fields where microlensing
events are found, it is necessary to use high angular resolution adaptive optics (AO) or 
HST observations to resolve the (possibly blended) lens and source stars
from other, unrelated stars. Unfortunately, this is not sufficient to establish a unique identification of the
lens (and planetary host) star \citep{moa310,highres}, so it is necessary to confirm that the host star is moving away
from the source star at the predicted rate \citep{ogle169,batistaogle169}. For that reason, we analyzed and confirmed in the previous section that the additional flux on top of the source is indeed the host lens star.

In order to obtain good sampling of light curve model parameters that are consistent with our photometric 
constraints and astrometry, we apply the following constraints, along with Galactic model constraints
when summing over our light curve modeling MCMC results to determine the final parameters.
The proper motion vectors of $\mu_{\rm rel,H}$ and the lens magnitudes are constrained according to Table \ref{tab-mparams} and \ref{tab-murel}. 
The $\mu_{\rm rel,H}$ constraints are applied to the Galactic model and the lens magnitude
constraints are applied when combining the MCMC light curve model results with the Galactic model.
The lens magnitude constraints require the use of a mass-luminosity relation. We built
an empirical mass luminosity relation following the method presented in \citet{bennett_moa291}.
This relation is a combination of mass-luminosity relations for different mass ranges. 
For $M_L \geq 0.66\,\msun$, $0.54\,\msun\geq M_L \geq 0.12\,\msun $, and 
$0.10 \,\msun \geq M_L \geq 0.07\,\msun$, we use the relations of \citet{henry93}, \citet{delfosse00},
and \citet{henry99}, respectively. In between these
mass ranges, we linearly interpolate between the two relations used on the
boundaries. That is, we interpolate between the \citet{henry93} and the \citet{delfosse00}
relations for $0.66\,\msun > M_L > 0.54\,\msun$, and we interpolate between the
\citet{delfosse00} and \citet{henry99} relations for $0.12\,\msun > M_L > 0.10\,\msun$. When using
these relations we assume a 0.05 magnitude uncertainty.

For the mass-luminosity relations, we must also consider the foreground extinction.
At a Galactic latitude of $ b = -4.7009^\circ$, most of the dust is likely to be in the foreground of
the lens unless it is very close to us. We quantify this with a relation relating the extinction 
of the foreground of the lens to the extinction in the foreground of the source.
Assuming a dust scale height of $h_{\rm dust} = 0.10\pm 0.02\,$kpc, we have
\begin{equation}
A_{i,L} = {1-e^{-|D_L(\sin b)/h_{\rm dust}|}\over 1-e^{-|D_S (\sin b)/h_{\rm dust}|}} A_{i,S} \ ,
\label{eq-A_L}
\end{equation}
where the index $i$ refers to the passband: $V$, $I$, $H$, or $K$.

These dereddened magnitudes can be used to determine the angular source radius,
$\theta_*$. With the source magnitudes that we have measured, the most precise determination
of $\theta_*$ comes from the $(V-I),I$ relation. We use
\begin{equation}
\log_{10}\left[2\theta_*/(1 {\rm mas})\right] = 0.501414 + 0.419685\,(V-I)_{s0} -0.2\,I_{s0} \ ,
\label{eq-thetaS}
\end{equation}
which comes from the \citet{boyajian14} analysis, but with the color range optimized for the 
needs of microlensing surveys \citep{aparna16}.

\begin{deluxetable}{cccc}
\tablecaption{Comparison of Measurement of Planetary System Parameters from 2018 vs 2005 Data\label{tab-params-histogram}}
\tablewidth{0pt}
\tablehead{\colhead{parameter}&\colhead{units}&\colhead{2018}&\colhead{2005}}
\startdata
%Angular Einstein Radius, $\theta_E$&mas&$0.322\pm 0.010$&0.303--0.342 \\
Geocentric lens-source relative proper motion, $\mu_{\rm rel, G}$&mas/yr&$3.31\pm 0.06$& $3.3\pm 0.4$\\
Host star mass, $M_{\rm host}$&${\msun}$&$0.56\pm 0.06$ &$0.63^{+0.07}_{-0.09}$\\
Planet mass, $m_p$&$M_{\rm Jup}$& $2.34\pm 0.43$& $2.6^{+0.8}_{-0.06}$\\
Host star-planet 3D separation, $a_{3D}$&AU&$4.0^{+2.1}_{-0.7}$&$4.3^{+2.5}_{-0.08}$\\
%Host star-planet 2D separation, $a_{\perp_{wide}}$&AU&$5.9 \pm 0.7$& 4.7--7.7\\
Lens distance, $D_L$&kpc &$5.26\pm 0.71$&$5.8^{+0.6}_{-0.7}$\\
%Lens magnitude, $K_L$& &$18.78\pm 0.11$& 18.57--18.99\\
%Lens magnitude, $H_L$& &$19.02\pm 0.12$& 18.81--18.99\\
%Lens magnitude, $I_L$& &$21.24\pm 0.10$& 21.07--21.53\\
%Lens magnitude, $V_L$& &$23.47\pm 0.21$& 23.08--24.01\\
\enddata
\end{deluxetable}

%\begin{figure}
%\epsscale{1.0}
%\plotone{mass_theta_400.pdf}
%\caption{The $H_L$ and $K_L$ brightness of lens yields the mass distance relations shown in the plot. These mass distance relations intersect the mass distance relation obtained from Einstein radius obtained from the light curve. The intersection of these different mass distance relations yields the mass and distance to the lens system. It is also to be noted that the mass distance relations from $H_L$ and $K_L$ brightnesses both independently intersect $\theta_E$ mass distance relation at the same place. This shows that the mass measurements made independently from two different passbands are consistent.}
%\label{fig-lens_theta}
%\end{figure}

\begin{figure}
\epsscale{1.0}
\plotone{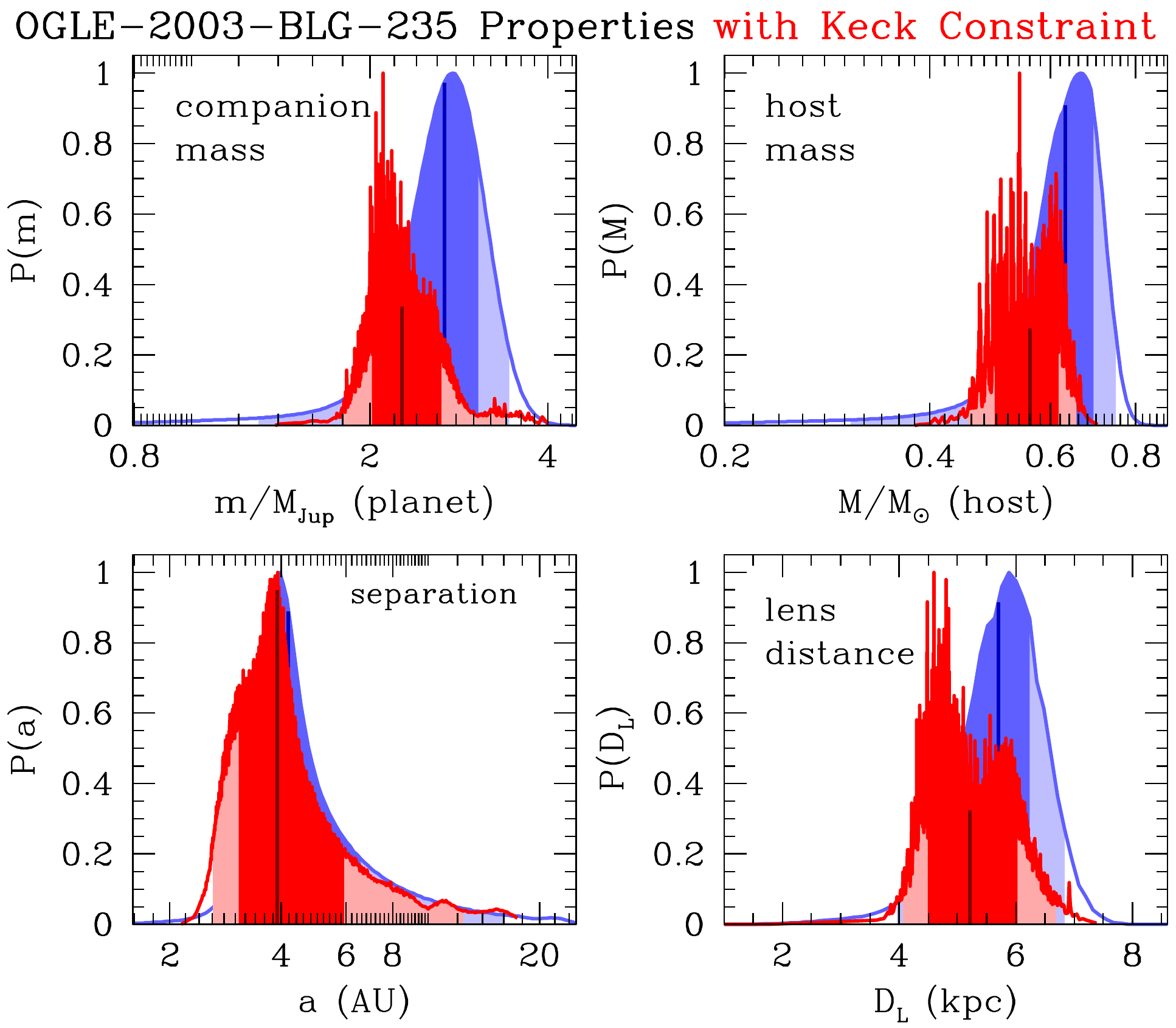}
\caption{The Bayesian posterior probability distributions for the planetary companion mass, host mass, 
their separation and the distance to the lens system are shown with the light curve + HST constraints in blue (based on \citet{bennett06}) 
and the constraints from our Keck follow-up observations in red.
The central 68.3$\%$ of the distributions are shaded in darker colors (dark red and dark blue) and the 
remaining central 95.4$\%$ of the distributions are shaded in lighter colors. The vertical black line marks 
the median of the probability distribution of the respective parameters. The older results are more smooth because they assume Gaussian uncertainties on the microlens model parameters, while new results include the full parameter distributions from the MCMC. This comparison shows the consistency of the 2018 mass measurement results from Keck to the 2005 mass measurement results with HST centroid shift measurements.}
\label{fig-lens-Keck}
\end{figure}

%\begin{figure}
%\epsscale{1.0}
%\plotone{lens_prop_ob03235_Keck_HST_Const_vs_old.pdf}
%\caption{The Bayesian posterior probability distributions for the planetary companion mass, host mass, 
%their separation and the distance to the lens system are shown with the light curve + HST constraints in blue (based on \citet{bennett06}) 
%and with the additional constraints from our Keck follow-up observations in red.
%The central 68.3$\%$ of the distributions are shaded in darker colors (dark red and dark blue) and the 
%remaining central 95.4$\%$ of the distributions are shaded in lighter colors. The vertical black line marks 
%the median of the probability distribution of the respective parameters. The older results are more smooth because they assume Gaussian uncertainties on the microlens model parameters, while new results include the full parameter distributions from the MCMC. The red graph shows a total result of the mass measurements with both the Keck and HST results together.}
%\label{fig-lens-Keck_HST}
%\end{figure}

We apply the $K$-band mass-luminosity relations to each of the models in our Markov Chains
using the mass determined by the first expression of equation~\ref{eq-m_thetaE}, using the $\theta_E$
value determined from $\theta_E = \theta_* t_E/t_*$, where $t_E$ and $t_*$ are light curve parameters
given in Table~\ref{tab-mparams}. We can then use the Keck $K$-band measurements of the
lens star brightness from Table~\ref{tab-Keck_phot} to constrain the lens brightness including both the
observational uncertainties listed in Table~\ref{tab-Keck_phot} and the 0.05 mag theoretical uncertainty
that we assume for our empirical $K$-band mass-luminosity relations. We solve the first expression of equation~\ref{eq-m_thetaE} along with the mass-luminosity relations. While solving, we assume the source distance $D_S  = 8.57\pm 1.42$ kpc. We solve this for every MCMC model and we sum over our MCMC results
using the Galactic model employed by \citet{bennett14} as a prior, weighted by the 
microlensing rate and the measured $\mubold_{\rm rel,H}$ values given in Table~\ref{tab-murel}. The results of our final sum over the Markov Chain light curve models are given in 
Table~\ref{tab-params-histogram} and Figure \ref{fig-lens-Keck}. This table gives the mean and RMS
uncertainty plus the central 95.4\% confidence interval range for each parameter except the 
3D separation, $a_{3D}$, where we give the median and the central 68.3\% confidence interval instead of the mean and RMS. The host mass is measured to be $M_{\rm host} = 0.56\pm 0.06\msun$, a K dwarf star, orbited by a 
super-Jupiter mass planet,
$M_{\rm Planet} = 2.34\pm 0.43 M_{\rm Jup}$. 

Figure \ref{fig-lens-Keck} and Table \ref{tab-params-histogram} show the comparison between mass measurements from HST and light curve model and the measurements only from Keck.   As we can see from Table \ref{tab-params-histogram}, the mass and distance measurements from both 2006 and 2018 data are consistent within $1\sigma$.

\section{Confirming the Centroid Shift Measurement}
\label{sec-centroidshift}
The lens was detected in \citet{bennett06} paper by measuring the centroid shifts between $B$, $V$ and $I$ passbands in HST data. This data was taken only 1.78 years after the peak of the event. Figure 2 of \citet{bennett06} shows the measurements of the centroid shift in different bands. The $x$ and $y$ axes are aligned with RA and Dec. These values are also given in the last two columns of Table \ref{tab-params_centroid}. 

As the lens separates out from the source star, it is detected in the 2018 Keck data. This detection is independent of the detection in 2006 HST data. To confirm the centroid shift measurements done in the previous paper, we predict the centroid shifts in $B-I$, $B-V$ and $V-I$ passbands. The $\mu_{rel,H}$ is $2.4674 \pm 0.0121$ mas/yr and $2.5670 \pm 0.0215$ mas/yr for East and North.  This gives the predicted separation at 1.78 years is $dE = 4.3920 \pm 0.0215$ mas and $dN = 4.5693 \pm 0.0532$ mas. Since we know the lens-source flux ratio in three passbands $B$, $V$ and $I$ from \citep{bennett06}, we predicted the centroids at a separation of $dE = 4.3920 \pm 0.0215$ mas and $dN = 4.5693 \pm 0.0532$ mas in $B$, $V$ and $I$ passbands. From this, we predicted the centroid shifts in $B-I$, $V-I$ and $B-V$ passbands. These predictions are shown in Table \ref{tab-params_centroid}.   
\begin{deluxetable}{ccccc}
\tablecaption{The Prediction of centroid shift from 2018 data vs actual 2006 Analysis\label{tab-params_centroid}}
\tablewidth{0pt}
\tablehead{\colhead{Passbands}&\multicolumn{2}{c}{Prediction based on 2018 data}&\multicolumn{2}{c}{HST Analysis from 2006}\\
&East&North&East&North}

\startdata
%Angular Einstein Radius, $\theta_E$&mas&$0.322\pm 0.010$&0.303--0.342 \\
$B-I$&$-0.52\pm 0.06$ &$-0.54 \pm 0.06$ &$-0.26\pm 0.36$ &$-0.62\pm0.24$ \\
$B-V$&$-0.14\pm 0.04$ &$-0.14 \pm 0.04$ &$-0.21\pm 0.29$ &$-0.61\pm0.23$ \\
$V-I$&$-0.38\pm 0.04$ &$-0.39 \pm 0.04$ &$-0.05\pm 0.27$ &$-0.01\pm0.22$ \\
\enddata
\end{deluxetable}

Note that the $B-I$ centroid shift matches pretty well, with both components, within 1$\sigma$. The East component of the $B-V$ centroid shift is also within 1$\sigma$, but North component is just outside 2$\sigma$. Both components of the $V-I$ centroid shift are within 1$\sigma$ and 2$\sigma$ of the prediction. Note that this imperfect match was obvious in the \citet{bennett06} paper. The $V$ centroid is too close to the $I$ centroid and too far from the $B$ centroid to be 100$\%$ consistent with normal stellar colors. 

This is the second event with the confirmation of color dependent centroid shift measurements. \citet{bennett20} has confirmed the mass measurement of \citet{dong-ob05071} which was observed by HST only 0.84 years after the peak of the event. 
\section{Discussion and Conclusions}
\label{sec-discussion}
OGLE-2003-BLG-235 was observed in 2018 to confirm the lens detection in the 2006 data that was done with centroid shift method. The color dependent centroid shift method will be one of the primary methods of mass measurement of exoplanets and their hosts with the Nancy Grace Roman Space Telescope. In this paper we wanted to demonstrate the feasibility of this method. We measured relative lens-source proper motion and the masses using 2018 Keck data, completely independent of the \citet{bennett06} paper, and found they are consistent. We predicted the centroid shifts based on the separation measured from 2018 data and found that they are consistent with the ones measured in the 2006 analysis. This shows that the color dependent centroid shift method provides reliable results.

   The color dependent centroid shift mass measurement is a powerful tool when determining masses of systems where the lens and source are merely separated. Currently this method is built and tested with 8-16 dithered images with HST and about 30 dithered images with Keck. In the Roman Galactic Exoplanet Survey, there will be six epochs spread out over 5 years. The maximum time separation between the first epoch and the last epoch will be about 4.5 years. Most of the events have an average 5 mas/year lens-source relative proper motion. Hence for most of the events, the lens-source separation will be about $\lesssim$ 22.5 mas. The pixel scale of the Roman telescope is 110 mas. Hence the lens-source separation will be around $\lesssim$0.2 parts of the pixel. Comparing to the 2006 studies, the lens-source separation was 0.16 parts of the pixel of HST ACS camera. In addition to that, Roman Galactic Exoplanet Survey will take a total of about 40000 images in the wide band and 800 dithered images in each additional passband (two - one red and one bluer than wide F146) in each of the 7 fields across six seasons. Each season consists of about 72 days of observations.  So each field will have $>$100 dithered images in different passbands in each season. The analysis of this huge number of dithers will provide us a very well sampled PSF and yield a significantly higher photometric and astrometric precision compared to a handful of HST images. This advantage will aid immensely in obtaining mass measurements of the lens (host and exoplanet) with the color dependent centroid shift method.  

The color dependent centroid shift method can also be utilized by joint Euclid-Roman survey \citep{euclid} to measure the masses of RGES microlensing planets. In this proposed multiband survey all the RGES fields will be observed by Euclid a few years before the launch of Roman. This brief, early Euclid survey will capture several future microlensing candidates where the lens and source are moving towards each other before the peak of the microlensing event. When Roman detects the microlensing planetary events, the Euclid multiband microlensing data will show the lens and source partially resolved and the color-dependent centroid shift method can be used to measure the host star mass. However, to note that Roman telescope itself will be able to measure masses of significant number of the microlensing planetary events using its high resolution capabilities. NASA requires RGES should be capable of determining the masses of, and distances to, host stars of 40$\%$ of the detected planets with a precision of 20$\%$ or better. This paper shows that it will be possible to detect masses and distances with a precision at par with the requirement.   
%However, the relatively large masses for the 
%planetary host stars MOA-2007-BLG-400L and MOA-2013-BLG-220L might be an indication that 
%M-dwarfs really are less likely to host gas giant planets (at a fixed mass ratio).
%On the other hand, \citet{bennett20} found that planet OGLE-2005-BLG-071Lb is a $q = 7\times 10^{-3}$ 
%planet orbiting an M-dwarf of $M = 0.43 \pm 0.04\msun$. This is a larger mass ratio, which might 
%suggest that this planet might have been formed by gravitational instability \citep{boss1997,cameron1978} instead of core
%accretion. It is also possible that the high metallicity of bulge stars plays a role \citep{bensby17},
%because high metalicity stars have been found to be more likely to host gas giant planets \citep{fischer05}.
%For events like OGLE-2005-BLG-071 \citep{bennett20}, OGLE-2005-BLG-169 \citep{batistaogle169}, 
%MOA-2007-BLG-400, and MOA-2013-BLG-220 \citep{van20}, that have lens stars are now resolved
%from their source stars, it should be possible to measure the metalicity. 
%Of course, we do need a larger sample of microlens planets with host star mass measurements to resolve 
%{\bf this issue}, and this is the goal of our NASA Keck KSMS and HST observing programs.

 This work made use of data from the Astro Data Lab at
NSF’s OIR Lab, which is operated by the Association
of Universities for Research in Astronomy (AURA), Inc.
under a cooperative agreement with the National Science Foundation. We also acknowledge the help of Dr. Peter Stetson on providing us with a feedback on our analysis of Keck data. The Keck Telescope observations and analysis
was supported by a NASA Keck PI Data Award 80NSSC18K0793. Data presented 
herein were obtained at the W. M. Keck Observatory from telescope time allocated to the National Aeronautics 
and Space Administration through the agency's scientific partnership with the California Institute of Technology 
and the University of California. The Observatory was made possible by the generous financial 
support of the W. M. Keck Foundation.
DPB, AB, NK, and SKT  were also supported by NASA through grant NASA-80NSSC18K0274 and 
by NASA award number 80GSFC17M0002.
This work was supported by the University of Tasmania through the UTAS Foundation and the endowed
Warren Chair in Astronomy and the ANR COLD-
WORLDS (ANR-18-CE31-0002). This research was
also supported in part by the Australian Government
through the Australian Research Council Discovery Pro-
gram (project number 200101909) grant awarded to
Cole and Beaulieu. Work by NK is supported by JSPS KAKENHI Grant Number
JP18J00897. J.R.L. acknowledges support by the National Science Foundation under Grant No. 1909641 and the Heising-Simons Foundation under Grant No. 2022-3542. AF's work was partly supported by JSPS KAKENHI Grant Number JP17H02871. 

%\begin{deluxetable}{cccc}
%\tablecaption{Measured Parameters from Single Star PSF fits \label{tab-single-fit}}
%\tablewidth{0pt}
%\tablehead{Passband&RA&Dec&Mag}
%\startdata
%HST $I$&&&\\
%HST $V$&&&\\
%Keck $K$&&&\\
%\enddata
%\end{deluxetable}
%\section{Bibliography}

\end{document}